\documentclass{camera}
\usepackage{graphicx}  % uncomment this if your want to insert figures

\begin{document}

%%%%%%%%%%%%%%%%%%%%%%%%%%%%%%%%%%%%%%%%%%%%%%%%%%%%%%%%
% The title, only the first letter capitalized; if you want to split it in
% two or more lines, put a \\ macro at each line break
% example: 
%   \title{Title: first line\\ second line}
%
\title{Determination of integral cross sections of $^3$H in Al foils monitors irradiated 
by protons with energies ranging from \\40 to 2600 MeV }

%%%%%%%%%%%%%%%%%%%%%%%%%%%%%%%%%%%%%%%%%%%%%%%%%%%%%%%%
% The author(s), separated by commas; do not put a
% comma before the last author, use instead the \and
% macro which produces a normal ``and'' in the
% caps/small caps context
%
\author{Yu.E. Titarenko$^1$, V.F. Batyaev$^1$, 
M.V. Chauzova$^1$, \\ I.A. Kashirin$^1$, 
S.V. Malinovskiy$^1$, K.V. Pavlov$^1$, V.I. Rogov$^1$, A.Yu. Titarenko$^1$,  V.M. Zhivun$^{1,2}$, S.G. Mashnik$^3$  \and \\ A.Yu. Stankovskiy$^4$}

%%%%%%%%%%%%%%%%%%%%%%%%%%%%%%%%%%%%%%%%%%%%%%%%%%%%%%%%
%
\organization{
$^1$NRC Kurchatov Institute $-$ \\ $“$Institute for Theoretical and Experimental Physics$”$, Moscow, Russia 

$^2$NRNU MEPhI (Moscow Engineering Physics Institute), Moscow, Russia 

$^3$Los Alamos National Laboratory, NM 87545, USA

$^4$SCK$\cdot$CEN, Boeretang, Mol, Belgium
}

%Affiliation}

\maketitle

\begin{abstract}
The results of $^3{H}$ production in $Al$ foil monitors ( $\sim$ 59 $mg/cm^2$ thickness) 
are presented. These foils have been irradiated in 15$\times$15 mm polyethylene bags of  $\sim$ 14 $mg/cm^2$ 
thickness together with foils of $Cr$ ($\sim$ 395 $mg/cm^2$ thickness) and  $^{56}{Fe}$ ($\sim$ 332 $mg/cm^2$ thickness) by protons of different energies in a range of 0.04 $ - $ 2.6 GeV. The diameters of all the
foils were 10.5 mm. 
The irradiations were carried 
out at the ITEP accelerator U$-$10 under the ISTC Project $\#$ 3266 in 2006$-$2009. $^3{H}$ has been extracted 
from $Al$ foils using an A307 Sample Oxidizer. An ultra low level liquid scintillation spectrometer 
Quantulus1220 was used to measure the $^3{H}$ $\beta-$spectra
and the SpectraDec software package was applied for spectra processing, deconvolution and $^3{H}$  activity determination.
The values of the $Al(p,x)^3{H}$ reaction cross sections obtained in these experiments are compared with 
data measured at other labs and with results of simulations by the MCNP6 radiation transport code using the
CEM03.03 event generator. 

\end{abstract}
%=========================================================================================================
\section{Introduction}
\begin{figure}[!ht]                                     %Fig{1}
\vspace*{-12 mm}
\hspace{-22 mm}
\includegraphics[width=17cm]{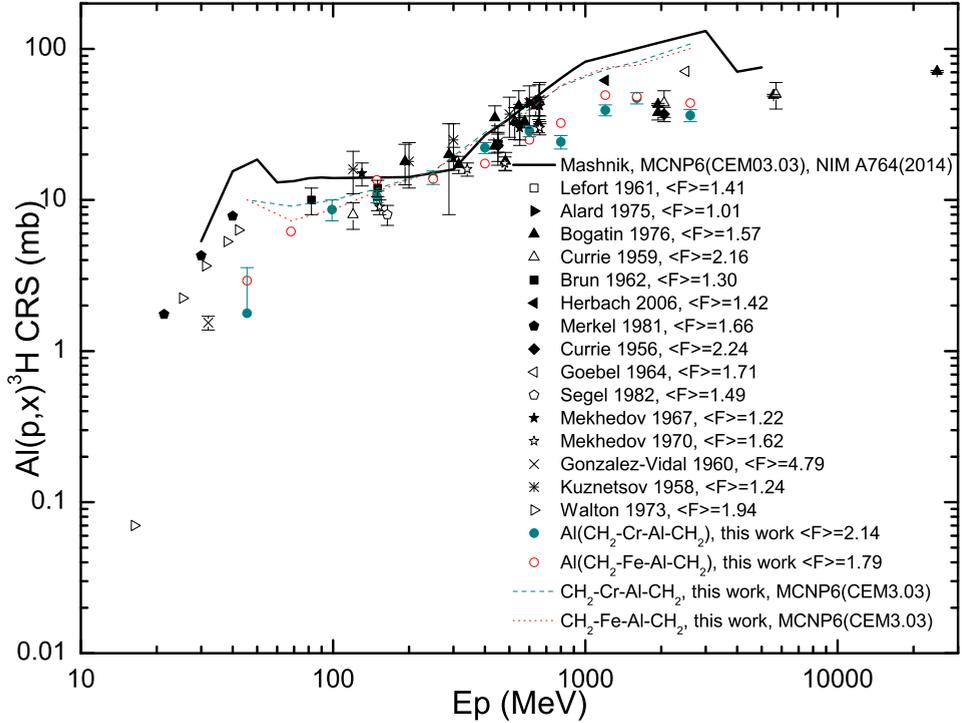}
\vspace*{-16 mm}
\caption{$Al$(p,x)$^3{H}$ reaction cross-sections measured in \cite{1,2}, simulated \cite{3} as well as obtained in this work.}
%\caption{Experimental \cite{1,2},  and simulated \cite{16} and got in this work cross-sections of the reaction $Al$(p,x)
%$^3{H}$.}
\label{Exfor}
\end{figure}

Tritium is a gaseous product of nuclear reactions, whose formation, in addition to the issues of radiation
production in specific parts of nuclear installations, causes additional environmental problems, associated
 with its high migration ability. This stimulates an interest in the study of the production cross-sections 
 of this nuclide in different structural materials of nuclear installations. A compilation of the experimental 
 values of the cross-sections of the reaction $Al$(p,x)$^3{H}$, taken from data libraries \cite{1,2} and together with values obtained in this work in comparison with the simulated data both in \cite{3} and in this work are presented in Fig. \ref{Exfor}.
 
A comparison of the simulated excitation functions of the $Al$(p,x)$^3{H}$ reaction with various experimental data sets reveals that the mean squared deviation factor $<F>$, often used to assess the predictive 
power of codes \cite{4}, is within the range from $\sim$ 1 to $\sim$ 5. 
In addition to the above, the spread of the experimental data is much wider than 
the experimental uncertainties.

%=========================================================================================================
\section{Justification of the experiments, simulation and results} 

During 2006 –- 2009, under the framework of the ISTC Project $\#$ 3266, the ITEP synchrotron was used to determine cross sections of residual nuclei production in $Cr$, $^{56}Fe$, and other thin samples induced by 0.04 -- 2.6 GeV protons. All the samples were placed inside polyethylene bags together with Al foils to monitor proton flux. Both the samples and foils were thin enough ($\ll$1$g/cm^2$) to get a negligible degradation of proton energy within each  ``sample-foil" sandwich. That is why the losses of residual nuclei heavier $^7{Be}$ could be neglected. The main project results obtained by $\gamma$--spectrometry of the irradiated samples can be found in \cite{4,5}.

\begin{figure}[!ht]                                           %Fig{2}
\vspace*{-10 mm}
\hspace{9 mm}
\includegraphics[width=11cm]{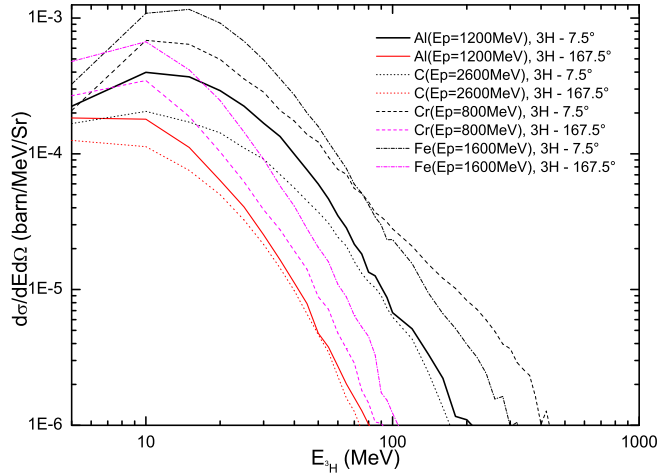}
\vspace*{-10 mm}
\caption{Energy spectra at different angles of $^3H$ nuclei released from the  $Al$, $C$, $Cr$ and $^{56}{Fe}$ nuclei
%targets
irradiated by 1200, 2600, 800 and 1600 MeV protons, accordingly.}
\label{SpectrAl}
\end{figure}

Subsequently, the idea arose to measure tritium inside the irradiated samples and foils. The event generator CEM03.03 inside the MCNP6.1.1 transport code was used to simulate the spectra of tritium nuclei produced by the protons and showed that a significant part of produced tritium has a range above the foils thickness (see Fig. \ref{SpectrAl}).

\vspace{- 1 mm}

 This fact leads to some redistribution of tritium produced in the polyethylene, samples, and monitor layers through the exchange of tritium nuclei. Therefore, the cross section results obtained in this work are not quite identical to the cross section for tritium production in Al presented in  \cite{1,2}; we identify our results as ``integral" cross sections. Nevertheless, as shown in this work, the obtained results can certainly be used to verify the nuclear models of radiation transport codes.

$^{3}{H}$ activity in $Al$ foil  was determined using a low-background spectrometric $\alpha$-, $\beta$-radiometer, Quantilus 1220, and 
a system of automatic sample preparation, A307 Sample Oxidizer.
An analysis of the measured $\beta$-spectra was carried out with the SpectraDec code
\cite{6}. The formula for determining the reaction rate and  cross-section production of $^{3}{H}$ are presented in \cite{5}.

All results were simulated using the CEM03.03 event-generator incorporated in the radiation 
transport code MCNP6.1.1 \cite{7}. The real parameters (thickness, diameter, density) 
of irradiated samples ($Al$, $Cr$, $^{56}{Fe}$) were modelled, including 
the thickness of polyethylene bags %(shells) 
in the appropriate layers-cells. 
%==========================================================================================================
\section{Conclusion and acknowledgements}
Simulated values of independent cross-sections of $^{3}{H}$ production in $Al$ \cite{5},
 together with the calculations of the integral cross-sections and measured values, are shown 
 in Fig. \ref{Exfor}. The values of $<F>$ for $Al$ from pair $Cr \leftrightarrow Al$ are equal to 2.14 and 1.79 from pair  $^{56}{Fe} \leftrightarrow Al$ (see details in \cite{5}).

The authors very much appreciate the support received from the ISTC projects, as well as
from the current pilot project of the National Research Center ``Kurchatov Institute". Part of
the work performed at LANL was carried out under the auspices of the National Nuclear
Security Administration of the U.S. Department of Energy. We thank Dr. Roger L. Martz for 
a very careful reading of the manuscript and useful suggestions on its improvement.

%%%%%%%%%%%%%%%%%%%%%%%%%%%%%%%%%%%%%%%%%%%%%%%%%%%%%%%%


\vspace*{-1mm}
\begin{thebibliography}{99}

%\vspace*{-2.0mm}
\bibitem {1} \emph{Experimental Nuclear Reaction Data (EXFOR) database}, http://www-nds.iaea.org/exfor/exfor.htm 
%Accession Nos. C0836009, D0628005, O0149002, O0304003, O0313002, C01160221, O0044025, O00460033, O0154024, O0305005, %O0342003, O1305002.

\vspace*{-2.0mm}
\bibitem {2}\emph {Production of Radionuclides at Intermediate Energies}, Ed. H.Schopper. Springer Verlag, Landolt-Bernstei.
%New Series, Group I, Vol 13, subvolumes 13a-13i (1991-1999), Ref. Mer81, Kuz58a, Wal73. 

\vspace*{-2.0mm}
\bibitem{3} S.G. Mashnik, L.M.Kerby, 
%MCNP6 fragmentation of light nuclei at intermediate energies
\emph{NIM A}, 764 (2014) 59–81.

\vspace*{-2.0mm}
\bibitem{4} Yu.E. Titarenko,  %V.F. Batyaev M.A. Butko,
et. al., % Verification of high-energy transport codes on the basis of activation data. 
\emph{Physical Review. C}, 84 (2011), 064612-12011.

\vspace*{-2.0mm}
\bibitem{5} Yu.E. Titarenko, %V.F. Batyaev, A.Yu. Titarenko, 
et. al., %Experimental and Theoretical Study of the Residual Nuclide Production in 40 – 2600 MeV Proton-Irradiated Thin %Targets of %ADS Structure Materials, 
\emph{IAEA Nuclear Data Section, INDC (CCP)-0453}, VIC, A-1400 Vienna, Austria, October 2011.


\vspace*{-2.0mm}
\bibitem{6} I.A. Kashirin, et al.,
%A.I. Ermakov, S.V. Malinovskiy, 
%et al.,%Liquid scintillation determination of low level components in complex mixtures of radionuclides. 
\emph{Applied Radiation and Isotopes}, 53, 303–308 (2000).

\vspace*{-2.0mm}
\bibitem{7} M.R. James, et. al.,
%D.B. Pelowitz, A. J. Fallgren,  et. al., 
%MCNP6TM User's Manual. Code Version 6.1.1 Beta, 
\emph{LA-CP-14-00745}, Los Alamos (2014).

\end{thebibliography}
\end{document}